\documentclass{vow2008}

\title{Data Reduction Pipeline for GTC/FRIDA}
\author{M.~C.~Eliche-Moral}
\author{N.~Cardiel}
\author{S.~Pascual} 
\author{J.~Gallego}
\affil{Dpto. de Astrof\'{\i}sica, Universidad Complutense de Madrid (Spain)}

\usepackage{graphicx}
\usepackage{multirow}
\begin{document}

\keywords{\LaTeX; Virtual Observatory; FRIDA; Integral Field Spectroscopy; IFU; Data Reduction}

\maketitle

\begin{abstract}
FRIDA (inFRared Imager and Dissector for the Adaptative optics system of the GTC) will be a NIR (1-2.5$\mu$m) imager and Integral Field Unit spectrograph to operate with the Adaptative Optics system of the 10.4 m GTC telescope. FRIDA will offer broad and narrow band diffraction-limited imaging and integral field spectroscopy at low, intermediate and high spectral resolution. The Extragalactic Astrophysics and Astronomical Instrumentation group of the Universidad Complutense de Madrid (GUAIX) is developing the Data Reduction Pipeline for FRIDA. Specific tools for converting output, reduced datacubes to the standard Euro3D FITS format will be developed, in order to allow users to exploit existing VO applications for analysis. FRIDA is to be commissioned on the telescope in 2011.
\end{abstract}

\vspace{-0.3cm}
\section{FRIDA Description}

FRIDA\footnote{More information on FRIDA at:\\
FRIDA Project URL:http://www.astroscu.unam.mx/ia\_cu/proyectos/frida/ \\
FRIDA Instrument URL: http://www.iac.es/project/frida/\\
FRIDA GUAIX URL: http://guaix.fis.ucm.es/instrumentation} will be a second generation instrument for GTC, and the first proposed for the Adaptative Optics (AO) system of the telescope \citep{Lopez07}. 
FRIDA is being designed and constructed by an international consortium, lead by the Instituto de Astronom\'{\i}a of the UNAM. It is an integral field spectrograph with imaging capabilities, optimized for the 1.1-2.4 $\mu$m range, and it will provide broad and narrow band imaging and Integral Field Spectroscopy (IFS) capabilities with low, intermediate and high spectral resolutions, and with different spatial resolutions in the NIR (see Table \ref{tab:table}). Its combination of high spectral and spatial resolutions will provide FRIDA with unique capabilities among other existing Integral Field Units (IFUs). FRIDA will be able to tackle a large number of astrophysical problems, from Solar system bodies to cosmological surveys. Its characteristics and its operating wavelength range will expose dusty environments, rich in stellar absorption and emission lines from many different species and gas phases, and sample kinematics and stellar or gaseous content of high redshift galaxies.

\begin{figure}
\centering
\includegraphics[width=0.95\linewidth]{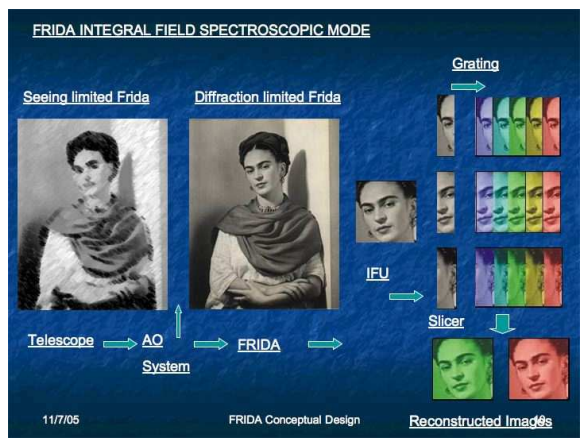}
\caption{FRIDA IFS mode diagram. Credits: A.~L\'{o}pez, S.~Cuevas.\label{fig:fig1}}
\end{figure}

FRIDA will perform imaging at the diffraction limit of the telescope, thanks to the correction performed to the wavefront by the GTC AO module. In the IFS mode, a monolithic image slicer composed by a set of mirrors will break down a small portion of the AO-corrected field into 30 slices, that will play the role of 1D spectroscopy slits. The light of each slice is dispersed by the spectrograph onto the detector. With these data, a 3D datacube can be built up containing a pseudo-monocromatic image of the IFU field-of-view (FoV) in each plane (see Fig.~\ref{fig:fig1} for a explanatory sketch of the datacube obtention). The FRIDA IFU design is based on FISICA IFU for FLAMINGOS-II \citep{Eikenberry04}.

\begin{figure}
\centering
\includegraphics[width=\linewidth]{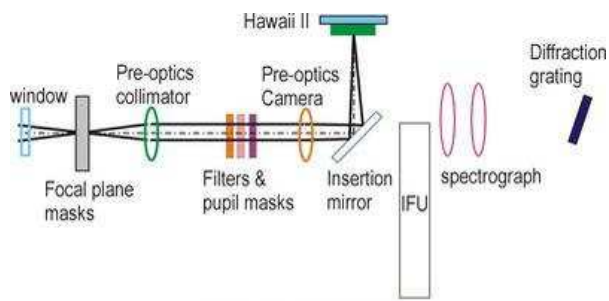}
\caption{FRIDA optical layout in imaging mode. Credits: A.~L\'{o}pez, S.~Cuevas.\label{fig:fig2}}
\end{figure}

\begin{figure}
\centering
\includegraphics[width=\linewidth]{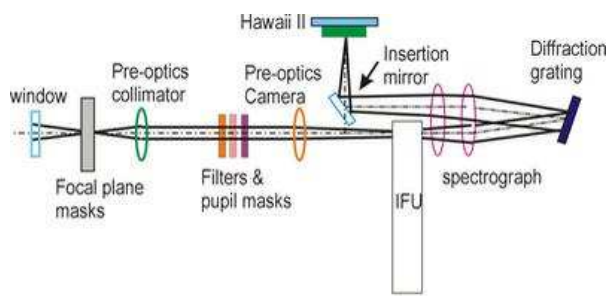}
\caption{FRIDA optical layout in IFS mode. Credits: A.~L\'{o}pez, S.~Cuevas. \label{fig:fig3}}
\end{figure}

Figures \ref{fig:fig2} and \ref{fig:fig3} show the light path through the different FRIDA optical components for the imaging and IFS modes, respectively \citep{Cuevas08}. In both cases, the light path starts at the cryostat window (on the left), coming directly from the GTC-AO module. It crosses the mask and filter wheels. The camera wheel selects the plate scale. In the IFS mode, the IFU dissects a small portion of the FoV into the 30 slices, and sends them to the grating carrousel for wavelength dispersion (Fig.~\ref{fig:fig3}). The dispersed light of each slice is finally focussed on the detector ("slitlet"). In the imaging mode, the beam avoids entering the IFU and a flat mirror replaces the grating in the spectrograph (Fig.~\ref{fig:fig2}).

\vspace{-0.3cm}
\section{FRIDA Data Reduction Pipeline}

GUAIX\footnote{GUAIX Home Page: http://guaix.fis.ucm.es/} is developing the Data Reduction Pipeline (DRP) for FRIDA. It will consist on robust software tools following certain quality assessment rules, specially optimized for FRIDA data. The DRP will perform two simultaneous and automatic data reduction processes in the telescope (per observing block or after the user's request): a quick data reduction for quick-look analysis, and a scientific-level reduction for archiving purposes in the GTC database. A stand-alone version is also under study. In order to avoid error correlation in those reduction steps that require interpolation or re-pixelation, error handling will be delayed during the reduction as far as possible. This ensures a realistic errors computation during the major part of the reduction \citep{Cardiel02}. Error flags will indicate the reliability of the computed errors to the user.

The DRP main characteristics are the following:
\begin{itemize} 
\item It is being coded following object-oriented structure. 
\item It will be composed of executable reduction recipes for generating completely-reduced data and final calibration products, for all the defined FRIDA operational modes.
\item Error handling and propagation throughout the data reduction are usually absent in other data reduction packages. FRIDA DRP will estimate them in parallel to the data reduction. Quality control checks will also be performed.
\item The final products of the DRP include: the processed calibrations, the reduced, re-constructed final datacubes, intermediate-reduced products, quality control flags. The pipeline final product will be a datacube per nominal pointing and observing block.
\end{itemize}

\begin{center}
\begin{table}
  \begin{center}
    \caption{Summary of FRIDA Characteristics}\vspace{1em}
    \begin{tabular}[h]{ll}
      \hline
	\multicolumn{2}{|c|}{General Characteristics} \\
      \hline                       
\multirow{2}{*}{Working location}     	&  Nasmyth-A GTC platform,\\
					& after the GTC-AO system \\
\multirow{2}{*}{Wavelength range}     	&  0.9-2.5 $\mu$m, optimized for \\
					& 1.1-2.4 $\mu$m\\
\multirow{2}{*}{Detector}		&  Rockwell Hawaii-II 2K$\times$2K, \\
			& HgCdTe \\
	\hline
	\multicolumn{2}{|c|}{Imaging Mode} \\
      	\hline
Imaging mode scales                	& 0.010, 0.020 \arcsec/pixel\\
FoV (arcsec$^2$)	& 20.48$\times$20.48, 40.96$\times$40.96\\
\multirow{2}{*}{Filters} 	& Broad band \emph{JHK}, and  \\
				& narrow band in 0.9-2.5 $\mu$m\\
	\hline
	\multicolumn{2}{|c|}{IFS Mode} \\
      	\hline
\multirow{2}{*}{IFS mode scales}  & 0.010 \arcsec/pixel $\times$ 0.020 \arcsec/slice,\\
				  & 0.020 \arcsec/pixel $\times$ 0.040 \arcsec/slice\\
Data format onto detector         & 30 slices $\times$ 66 pixels/slice\\
FoV (arcsec$^2$)  & 0.66$\times$0.60, 1.32$\times$1.20 \\
Spectral resolutions  	& R$\sim$1500, 4000, and 30,000\\
      \hline \\
      \end{tabular}
    \label{tab:table}
  \end{center}
\end{table}
\end{center}

\vspace{-0.7cm}
\section*{Acknowledgments}

We acknowledge support from the Madrid Regional Government through the ASTRID Project (S0505/ESP-0361), for development and exploitation of astronomical instrumentation (http://www.astrid-cm.org/). Partially funded by the Spanish MEC under the Consolider-Ingenio 2010 Program grant CSD2006-00070: "First Science with the GTC" (http://www.iac.es/consolider-ingenio-gtc/).

\end{document}